\documentclass[twocolumn,showpacs,pra,eqsecnum]{revtex4}

\usepackage{xspace,psfig,epsfig,amsfonts,amssymb,color}

\begin{document}

\newcommand{\beq}{\begin{equation}} \newcommand{\eeq}{\end{equation}}
\newcommand{\bqa}{\begin{eqnarray}} \newcommand{\eqa}{\end{eqnarray}}
\newcommand{\nn}{\nonumber} \newcommand{\nl}[1]{\nn \\ && {#1}\,}
\newcommand{\erf}[1]{Eq.~(\ref{#1})}
\newcommand{\erfs}[2]{Eqs.~(\ref{#1})--(\ref{#2})}
\newcommand{\crf}[1]{Ref.~\cite{#1}}
\newcommand{\dg}{^\dagger}
\newcommand{\rt}[1]{\sqrt{#1}\,}
\newcommand{\smallfrac}[2]{\mbox{$\frac{#1}{#2}$}}
\newcommand{\half}{\smallfrac{1}{2}}
\newcommand{\bra}[1]{\langle{#1}|} \newcommand{\ket}[1]{|{#1}\rangle}
\newcommand{\ip}[2]{\langle{#1}|{#2}\rangle}
\newcommand{\sch}{Schr\"odinger} \newcommand{\hei}{Heisenberg}
\newcommand{\bl}{{\bigl(}} \newcommand{\br}{{\bigr)}}
\newcommand{\ito}{It\^o} \newcommand{\str}{Stratonovich}
\newcommand{\sq}[1]{\left[ {#1} \right]}
\newcommand{\cu}[1]{\left\{ {#1} \right\}}
\newcommand{\ro}[1]{\left( {#1} \right)}
\newcommand{\an}[1]{\left\langle{#1}\right\rangle}
\newcommand{\implies}{\Longrightarrow}
\newcommand{\tr}[1]{{\rm Tr}\sq{ {#1} }}
\newcommand{\del}{\nabla} \newcommand{\du}{\partial}
\newcommand{\dbd}[1]{{\partial}/{\partial {#1}}}
\newcommand{\tp}{^{\top}}
\newcommand{\tbt}[4]{\left( \begin{array}{cc} {#1}& {#2} \\ {#3}&{#4} \end{array}\right)}

\title{Entanglement distribution by an arbitrarily inept delivery service}
\author{S. J. Jones, H. M. Wiseman, and D. T. Pope}
\affiliation{Centre for Quantum Computer Technology, Centre for
  Quantum Dynamics, School of Science, Griffith University, Brisbane,
  4111 Australia}
\date{June 24, 2005}

\begin{abstract}
We consider the scenario where a company $C$ manufactures in bulk
 pure entangled pairs of particles, each pair intended for a distinct pair of distant
 customers. Unfortunately, its delivery service is
inept -- the probability that any given customer pair receives its
intended particles is
 $S \in (0,1)$, and the customers cannot detect whether an error has occurred.
 Remarkably, no matter how small $S$ is, it is still possible for $C$ to distribute
 entanglement by starting with {\em non-maximally entangled} pairs. We determine
 the maximum entanglement distributable for a given $S$, and also
 determine the ability of the parties to perform
 nonlocal tasks with the qubits they receive.
\end{abstract}


\maketitle

\section{Introduction} Entanglement is an important resource in developing new
quantum technologies.  It is essential for many quantum
information processing (QIP) tasks such as quantum computation,
quantum teleportation and super-dense coding \cite{Nielsen,
Preskill}. At least for pure states and two parties, the
mathematical characterization of entanglement is well established.
However, it is only recently that researchers have begun to ask
questions about \emph{constrained} entanglement. That is,
entanglement that seems to exist in a system according to our
conventional description of its quantum states may not exist (or
at least may be of a different nature) because of constraints that
limit our ability to process the quantum information in the
system. In the case that the constraint (either fundamental or
practical) can be expressed as a super-selection rule (SSR), the
modification to our notions of entanglement are now well studied
\cite{VerCir03,WisVac03,BarWis03,BarDohSpeWis05,Schuch04,
Kitaev04}.

In this paper we consider a different sort of constraint that can
be understood from the following scenario. A company $C$
manufactures a pure entangled pair of particles for a pair of
customers, and delivers one particle to each customer. For reasons
of economy it manufactures these particle pairs in bulk, and
delivers to a large number of customer pairs. Unfortunately, its
delivery service is inept, meaning that the  probability that any
given customer pair receives particles that were manufactured as
an entangled pair is $S \in (0,1)$. Moreover, the customers cannot
detect whether an error has occurred. For example, the particles
may be indistinguishable. Thus the constraint is caused by a loss
of classical information, as in Ref.~\cite{Eisert} (the
differences with their situation will be expanded upon later). We
prove the surprising result that, no matter how small the success
probability $S$, $C$ can still distribute entanglement. We also
discuss how this scenario, although it might sound artificial, may
actually be relevant to the production of entangled photon pairs.

This paper is organized as follows. In Sec. II we present the
scenario in more detail. We then prove in Sec. III that
entanglement distribution is always possible, and follow this up
by determining the optimal entanglement distribution protocol in
Sec IV. Since distillation may not always be practical, in Sec. V
we study the ability of the mixed-up states to demonstrate
nonlocality without distillation. Section VI concludes with a
discussion of the practical implications of our theory in quantum
optics and ensemble quantum information processing.

\section{Scenario} 
\begin{figure}
\includegraphics[width=3in]{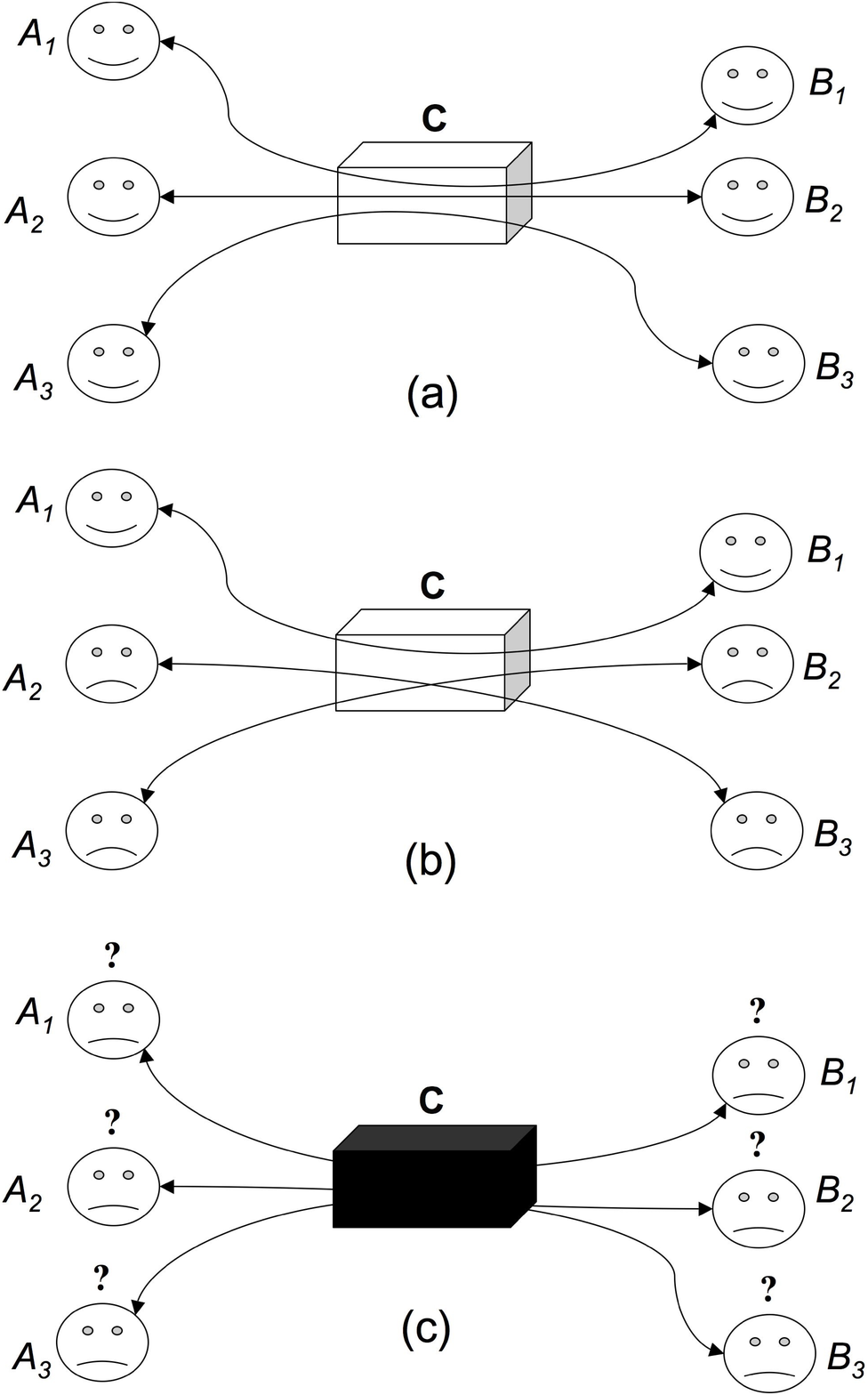}
\caption{In (a) we have an ideal case where $S=1$ and all the
customers receive their intended qubits. Situation (b) would be
typical if $S=1/3$, only one pair out of three receives their
intended qubits. In reality (c) would occur, the customers know
that $S<1$ and are unsure as to whether they have received an
entangled pair.} \label{fig:Fig1}
\end{figure}

We can now be more detailed about the scenario
outlined above. First, we assume that company $C$ manufactures
pure, identical, entangled pairs of {\em qubits}. Ideally, it
delivers one qubit from each pair  to remote customers, say, $A_1$
in Albuquerque, $A_2$ in Ajax, $A_3$ in Athens {\em etc.}, and the
other qubit from each pair to their respective partners $B_1$ in
Belfast, $B_2$ in Brisbane, $B_3$ in Berlin {\em etc}. This is
illustrated in Fig.~\ref{fig:Fig1}(a), and would allow any pair of
customers, $A_i$ and $B_i$, to undertake nonlocal quantum
information tasks such as teleportation \cite{Bennett93} or
violating a Bell inequality \cite{Bell}.

Now we will define the success probability $S$ of the service to
be the probability that any pair $(A_i,B_i)$  receives their
intended qubits in a pure entangled state. We will assume that $S$
is independent of $i$, and also that a failure (which occurs with
probability $1-S$) means that  one of $A_i$ and $B_i$ receive a
qubit other than the one intended for them, so that the qubits are
{\em uncorrelated}. For example, for three pairs of customers if
$S=1/3$ then a typical outcome would be the case shown in
Fig.~\ref{fig:Fig1}(b), where only one successful delivery takes
place.

In this case illustrated in Fig.~\ref{fig:Fig1}(b), one pair of
customers is still happy, as it knows it has received a pure
entangled pair. But what if the customers know that the delivery
service is inept, so that $S<1$, but  have no way of knowing
whether they actually received their intended qubits?  This is
illustrated by Fig.~\ref{fig:Fig1} (c), where the customers know
that $S=1/3$ as in Fig.~\ref{fig:Fig1} (b), but now all are unsure
as to whether they have an entangled pair.

Note that we need not assume that the qubits intended for the
customers $\{A_i\}$ are mixed up only amongst themselves; they
could be mixed up with the qubits for the $\{B_i\}$. However, for
simplicity, let us make that assumption. If, for the moment, we
further assume that the delivery service is otherwise completely
inept, so that for $N$ pairs of customers $S=1/N$, then our
scenario is close to that considered by Eisert {\em et al.}
\cite{Eisert} (which was formalized as a $S_N$-SSR in
Ref.~\cite{BarWis03}). The key difference is that (in our
language) Eisert {\em et al.} allow for all $N$ of the $A_i$s to
get together to do joint operations on their qubits, and likewise
the $B_i$s. In our scenario this is inappropriate as the customers
are distant from, and unaware of, each other. Thus, except between
members of a pair, even classical communication is forbidden.

The above considerations establish that in our scenario, each pair
of customers can be treated independently of each other pair.
Thus, to determine whether any entanglement has been delivered to
a particular pair of customers, whom we will call $A$ and $B$, we
consider their {\em state of knowledge}. Say the pure entangled
states prepared by $C$ are represented by $\rho_{AB}$. Then $A$
and $B$ know that they received this state with probability $S$
and that they received uncorrelated qubits (derived by throwing
away their entangled partner) with probability $1-S$. Thus, their
state of knowledge is \beq
 {\cal M}[\rho_{AB}] \equiv S\rho_{AB} +(1-S) {\rm
Tr_B}[\rho_{AB}]\otimes {\rm Tr_A}[\rho_{AB}]
\label{NonLinearMap}.
\end{equation}
Here ${\cal M}$ is a {\em nonlinear} map that describes the
mixing-up caused by the inept delivery service. Nonlinear maps of
this sort, with $S=0$, have been studied before as a ``universal
disentanglement machine" \cite{Terno}. For a single system (as
opposed to an ensemble of identical systems) this map is
unphysical \cite{Terno}; physical universal disentanglers which
operate imperfectly or probabilistically have also been considered
\cite{Buzek}.

 Na\"ively, we might expect that under almost
complete mixing-up (that is, $S\ll 1$) the customers would lose
their entanglement. After all, how could two parties have nonlocal
correlations if their subsystems almost certainly never interacted
with each other in the past? Surprisingly, this is not the case,
as we now show.

\section{Entanglement Distribution is Always Possible}
To carry out many QIP tasks it is useful to work with maximally
entangled states such as Bell pairs.  However, Bell pairs have
actually been found to be the most fragile states under the
influence of certain types of noise \cite{Fragilityexamples}. This
suggests that Bell states may be fragile under the influence of
mixing-up as we have defined it. Therefore, to distribute
entanglement when mixing up will occur, it may be more useful to
prepare entangled states of the form $\rho_{AB} =
\ket{\psi_a}\bra{\psi_a}$, where
\begin{equation}
\ket{\psi_a} = a \ket{0,0} + \sqrt{1-a^2}\ket{1,1},
\label{InitialState}
\end{equation}
with $a \in [0,1]$.  If $a=1 / \sqrt{2}$ then $\ket{\psi}$ is a
maximally entangled Bell state. If $a=0$ or $1$ then $\ket{\psi}$
is unentangled. Note that this state is symmetric under
interchange of $A$ and $B$, so here we can allow for the delivery
service to mix up qubits intended for the customers $\{A_i\}$ with
those intended for the customers $\{B_i\}$.

To gain an insight into which states will be most robust we
calculate the fidelity \cite{Jozsa}, $f$, of the state under
${\cal M}$: \beq f =\bra{\psi_a}{\cal M}\bigl[
\ket{\psi_a}\bra{\psi_a} \bigr]\ket{\psi_a} = 1 -3a^2(1-S)(1-a^2).
\label{Fidelity} \eeq The minimum of \erf{Fidelity} is at
$a=1/\sqrt{2}$, so Bell states are also the most fragile states
under the influence of mixing-up. By contrast, the fidelity of
\emph{non-maximally} entangled states with $a \ll 1$ remains high
even when a large amount of mixing-up occurs (i.e. $S\ll1$). This
suggests that  such states may be best for distributing
entanglement under the map ${\cal M}$.

\begin{figure}
\includegraphics[width=3in]{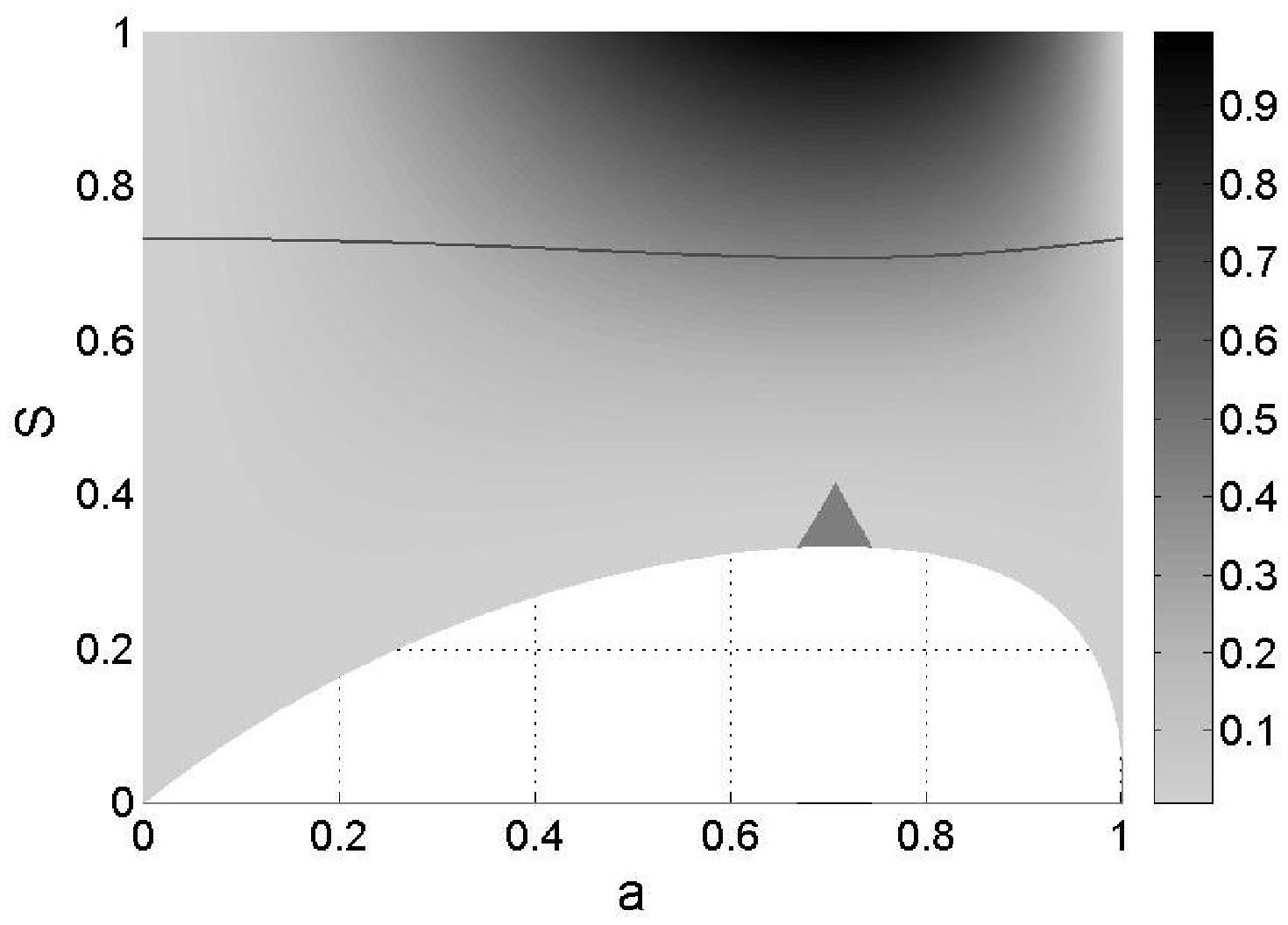}
\caption{Diagram to illustrate conditions under which $C$ can
distribute `useful' entanglement.  The colorbar gives an
indication of the amount of entanglement at each point on the
graph.  The white region is defined by
\erf{EntanglementDistributed} and demonstrates where no
entanglement is distributed. Above the dark line
\erf{CHSHViolationRequirement} is satisfied. Thus above this line
undistilled pairs can be used to violate the CHSH-Bell inequality.
In the gray regions below this line, entangled pairs are
distributed. However it is unclear if these undistilled pairs may
be used for nonlocal tasks.  In the region marked by a gray
triangle a LHVT for non-sequential POVM measurements exist (this
coloring is separate from the gray scale).} \label{fig:Fig3}
\end{figure}

To determine this, we calculate the concurrence $\mathcal{C}$
\cite{Wootters98}  of ${\cal M}[\ket{\psi_a}\bra{\psi_a}]$:
\begin{equation}
\mathcal{C} =2Sa\sqrt{1-a^2}-2(1-S)a^2(1-a^2).
\label{Concurrence}\end{equation} When $\mathcal{C}>0$ we know
that entanglement has survived despite the mixing-up of the
qubits. In particular, $\mathcal{C}>0$  iff (if and only if)
\begin{equation}
S > \frac{a(1-a^2)}{\sqrt{1-a^2}+a-a^2}.
\label{EntanglementDistributed}
\end{equation}
It is easy to verify that for any $S$, there is a range of $a$
values that satisfy this inequality. For $S \ll 1$,
\erf{EntanglementDistributed} is satisfied if $a < S$, as  can be
seen from  the bottom left-hand corner of Fig.~\ref{fig:Fig3}.
Thus, entanglement can \emph{always} be distributed by $C$ by
preparing an initial state $\ket{\psi_a}$ with sufficiently {\em
little} entanglement (the concurrence of $\ket{\psi_a}$ scales as
$2a$ for $a\ll1$). By contrast,  ${\cal C} = 0$ for Bell pairs
when $S<1/3$, due to their fragility.

\section{Optimal Entanglement Distribution}
\begin{figure}
\includegraphics[width=3in]{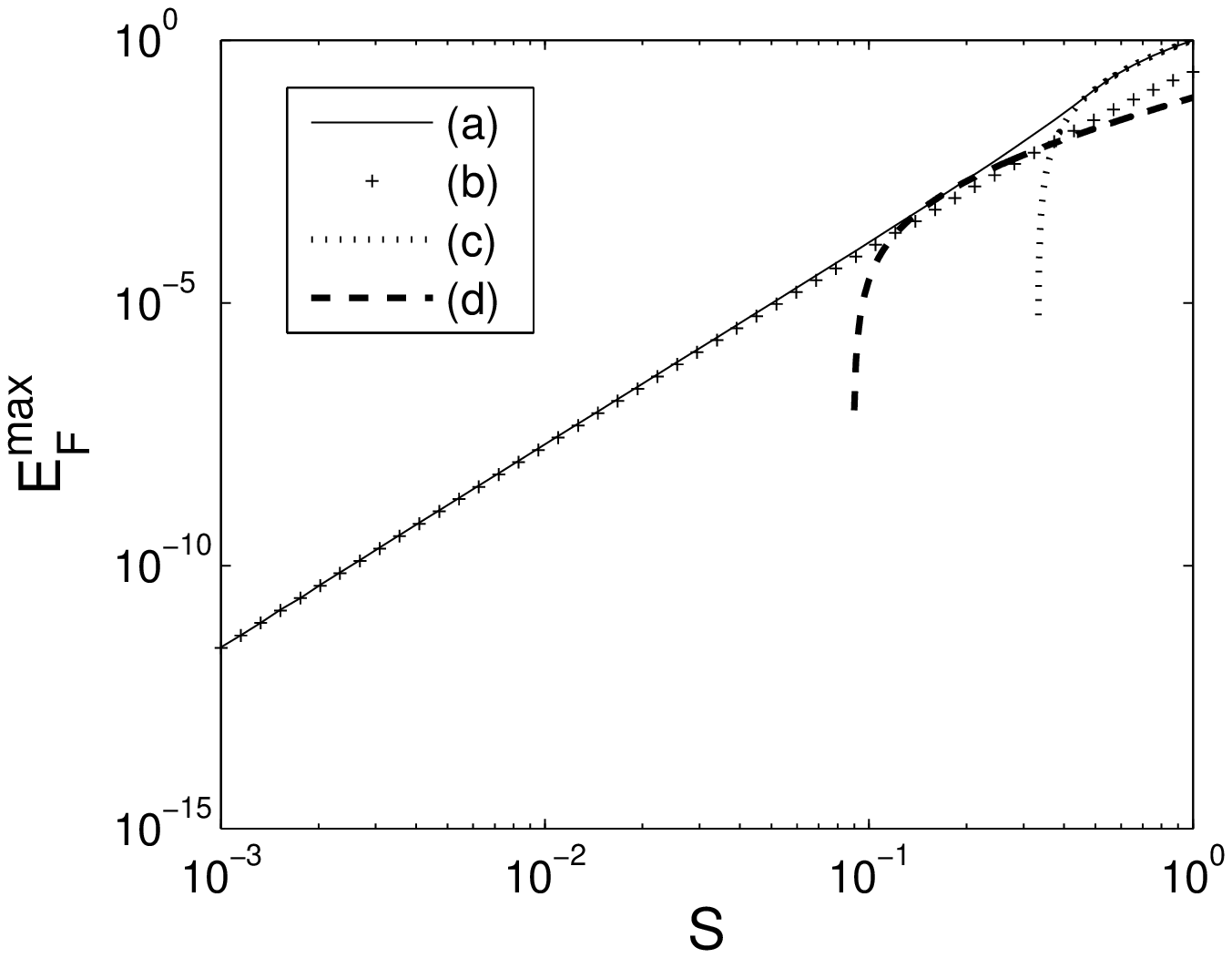}
\caption{Diagram of entanglement distributed for given success
probabilities. Curve (a) shows the maximum entanglement
distributed. The curve (b) is an approximate expression for the
maximum entanglement for $S$ given by \erf{Emax}. The plots (c)
and (d) show the entanglement for the specific cases when $C$
distributes initial states (\ref{InitialState}) defined by
$a=1/\sqrt{2}$ and $a=0.1$ respectively.} \label{fig:Fig2}
\end{figure}

We have shown that although inept delivery reduces the amount of
entanglement that can be accessed, entanglement can still be
distributed for arbitrarily low success probabilities. However, if
this entanglement is to be used as a resource for QIP we need to
know exactly how much entanglement can be accessed.  In the regime
$S\ll1$, \erf{Concurrence} becomes
\begin{equation}
\mathcal{C} \approx 2aS-2a^2. \label{ConcSmalla}
\end{equation}
The entanglement \footnote{Since $E_F$ quantifies the resources
needed to {\em create} a given entangled state, we would rather
use $E_D$, the \emph{distillable} entanglement. However,  the
latter cannot be computed, so we use $E_F$, which is an upper
bound on $E_D$ \cite{Bennett96}.} of formation $E_F$
\cite{Bennett96} for a pair of qubits is a monotonic function of
the concurrence, as shown by Wootters \cite{Wootters98}.   Thus
the maximum of $\mathcal{C}$ corresponds to the maximum of $E_F$.
It is clear from \erf{ConcSmalla} that $\mathcal{C}$ (and thus
$E_F$) is maximized for $a= S/2$.  Thus, in the $S\ll1$ regime we
have an approximate analytical expression for the maximum amount
of entanglement that can be distributed
\begin{equation}
E_F^{\rm max} \approx
\frac{S^4}{4}\left[\log_2\left(\frac{1}{S}\right)+1+\frac{1}{4\ln(2)}\right]
\label{Emax}.
\end{equation}
A comparison of \erf{Emax} and results for the maximum
entanglement of formation which can be distributed (determined
numerically) is shown in Fig.~\ref{fig:Fig2}. Thus we can say
that, given a success probability $S$, to distill a single Bell
pair requires delivery of \emph{at least} of order
$4/S^4\log_2{S^{-1}}$ non-maximally entangled pairs.

\section{Nonlocality Without Distillation}
In some instances it may be impractical to perform distillation
protocols to retrieve Bell pairs.  Another interesting question is
therefore whether or not the entanglement present in the initial
pairs \emph{without} distillation could be useful for some
nonlocal task. One test of this is to see if a Bell inequality
violation can be demonstrated.

While entanglement can be distributed for arbitrarily low $S$,
this entanglement {\em cannot} be used to violate a Bell
inequality at low $S$. For example, it is simple to show that
 the Clauser-Horne-Shimony-Holt (CHSH) Bell inequality
\cite{CHSH} is violated iff
\begin{equation}
S > \frac{1}{4(4a^4-4a^2-1)(4a^4-4a^2+1-\sqrt{4a^4-4a^2+3})}.
\label{CHSHViolationRequirement}
\end{equation}
 For Bell states this requires
that $S>1/\sqrt{2}$, and for $a\neq1/\sqrt{2}$ the requirement is
more strict, as shown by the line in Fig.~\ref{fig:Fig3}. It can
thus be seen that relatively high success probabilities are
required in order for the undistilled pairs to violate the CHSH
inequality .

In the gray region of Fig.~\ref{fig:Fig3} below this line it is
unclear whether or not the pairs can be used for nonlocal tasks
without distillation.  It may be possible to find other Bell
inequalities which would allow a wider class of the undistilled
pairs to be used to demonstrate nonlocality.  However, we have
shown that for at least part of this region there exists a
local-hidden-variable theory (LHVT)  for any non-sequential POV
measurement \cite{Werner89, Barrett02}. That is, each party can
make an arbitrary measurement, but is not allowed to communicate
the result to the other party.  The proof is straightforward and
follows work of Barrett \cite{Barrett02}. He showed a LHVT exists
under these conditions for the non-separable state (a Werner
state)
\begin{equation}
\rho_{\rm B} = \frac{5}{12}\ket{\psi}\bra{\psi}
+\frac{7}{12}\left(\frac{I}{2}\otimes\frac{I}{2}\right),
\label{BarrettState}
\end{equation}
where $\ket{\psi}$ is a Bell state ($a=1/\sqrt{2}$), and $I$ is
the $2\times 2$ identity matrix.

We show that for some values of $a$, ${\cal
M}\left[\ket{\psi_a}\bra{\psi_a}\right]$ can be written as a
convex combination of \erf{BarrettState} and a separable state:
\begin{equation}
\rho = {\cal M}\left[\ket{\psi}\bra{\psi}\right] = c \rho_{\rm B}
+ (1-c) \rho_{\rm SEP}, \label{LHVTState}
\end{equation}
where $0 < c < 1$. For such states $\rho$ it is clear that a LHVT
exists, since such a theory always exists for separable states. By
rearranging \erf{LHVTState}, we have
\begin{equation}
\rho_{\rm SEP}=(\rho - \rho_{\rm B})/(1-c).
\end{equation}
Thus, in order for our LHVT to work, $\rho_{\rm SEP}$ should be a
valid separable density matrix.
For simplicity, we considered only $\rho_{\rm SEP}$ states that
were diagonal.

Under these conditions we identified a finite region of $S$--$a$
space for which a LHVT can be found for entangled states given by
\erf{NonLinearMap}. While the above reasoning is straightforward,
actually calculating the region is more challenging. It requires
$\rho_{\rm SEP}$ to have all non-negative eigenvalues and trace
equal to one. Two of these eigenvalues are found to always be
positive. Thus, the region shown for which the LHVT exists is
bounded by three conditions; that the remaining two eigenvalues
are non-negative (left and right boundaries), and that
entanglement is distributed (lower boundary).  This is
demonstrated in Fig.~\ref{fig:Fig3}.

\section{Discussion}
We have considered the problem of the distribution of entanglement
to two parties, $A$ and $B$, by an inept
 delivery service that has only a probability $S$ of
successfully delivering the intended qubits. (If it fails, then
$A$ or $B$ receive a qubit intended for some other party). We have
shown that no matter how small $S$ is, entanglement can still be
distributed if the source supplies identical pure qubit pairs that
are {\em non-maximally entangled}.

It might be thought that this is an artificial problem, but that
is not necessarily the case. Consider the production by parametric
downconversion of entangled pairs of photons~\cite{Kwi01}. In some
experiments, these photons may enter many transverse spatial
modes, but the detectors used in the experiment may not resolve
these modes (i.e. all modes enter the detector). In the limit of
small flux, each photon detection is well separated in time and so
it is easy to tell that a coincidence count (counts at the two
detectors within some
 time-window) corresponds to an entangled pair. But
in the limit of high flux, a coincidence thus defined might
actually be due to photons from two different entangled pairs
(with different transverse spatial modes). That is, there is only
a probability $S$ (that could be quite low) that a pair of photons
identified by their arrival time is actually a pair produced by
downconversion. Thus the scenario we have
 described in terms of an inept delivery service could arise naturally
 in a quantum optical experiment. The solution to the mixing-up problem that
 we have identified might be of practical use in these or similar
  quantum information experiments.

  We conclude by returning to the situation of Eisert {\em et al.} \cite{Eisert},
  as mentioned earlier, where there are $N$ pairs of customers and $S=1/N$.
  In that case, the result we have obtained for $E_F^{\rm max}$ (\ref{Emax}), multiplied by
  $N$, is (ignoring the distinction between $E_F$ and $E_D$) a {\em lower bound}
  on the amount of entanglement distillable in the situation of Eisert {\em et al.},
 where the $A$ customers can all get together and likewise the $B$ customers.
 This might seem useless, since the distillable entanglement was calculated exactly
 in Ref.~\cite{Eisert} (see also Ref.~\cite{BarWis03}). However, our $E_F^{\rm max}\times N$
 is also a lower bound on the entanglement distillable by $\{A_i\}$ and  $\{B_i\}$ under
 the stronger constraint that they can only implement operations on individual qubits,
 not collective (entangling) operations on all $N$ qubits (as allowed in Refs. \cite{Eisert,BarWis03}). The relevance of this stronger
 constraint to NMR quantum information processing will be discussed in a future work.

\begin{acknowledgments}

  This work was supported by the Australian Research Council and
  the State of Queensland. We thank N. Gisin for pointing out the application
  of our work in quantum optics, and acknowledge discussions
  with S. D. Bartlett and M. A. Nielsen.

\end{acknowledgments}


\begin{thebibliography}{99}

\bibitem{Nielsen}
M. A. Nielsen and I. L. Chuang, \emph{Quantum Computation and
Quantum Information}, Cambridge University Press, (2000).

\bibitem{Preskill}
J. Preskill, Physics 229:Advanced Mathematical Methods of Physics
- Quantum Computation and Information. California Institute of
Technology (1998). URL:
http://wwww.theory.caltech.edu/people/preskill/ph229

\bibitem {VerCir03} F. Verstraete and J. I. Cirac,
Phys. Rev. Lett. {\bf 91}, 010404 (2003).

 \bibitem{WisVac03}
 H. M. Wiseman and J. A. Vaccaro,
Phys. Rev. Lett. {\bf 91}, 097902 (2003).

 \bibitem{BarWis03}
  S. D. Bartlett and H. M. Wiseman,
Phys. Rev. Lett. {\bf 91}, 097903 (2003).

\bibitem{BarDohSpeWis05}
S. D. Bartlett, A. C. Doherty, R.W. Spekkens, and H. M. Wiseman
``Entanglement under restricted operations: an analogy to mixed
state entanglement'', quant-ph/0412158.

\bibitem{Schuch04}
N. Schuch, F. Verstraete, and J. I. Cirac, Phys. Rev. A {\bf 70},
042310 (2004).

\bibitem{Kitaev04}
A. Kitaev, D. Mayers, and J. Preskill, Phys. Rev. A {\bf 69},
052326 (2004).

\bibitem{Eisert}
J. Eisert, T. Felbinger, P. Papadopoulos, M. B. Plenio, and M.
Wilkens, Phys. Rev. Lett. {\bf 84}, 1611 (2000).

\bibitem{Bennett93} C. H. Bennett, G. Brassard, C. Cr\'epeau, R. Jozsa, A. Peres, and W. K. Wootters,
Phys. Rev. Lett. {\bf 70}, 1895 (1993).

\bibitem{Bell} J. S. Bell, Physics {\bf 1}, 195 (1964).

\bibitem{Terno} D. R. Terno, Phys. Rev. A {\bf 59}, 3320 (1998).

\bibitem{Buzek} V. Bu\v{z}ek and M. Hillery, Phys. Rev. A {\bf 62},
052303 (2000).

\bibitem{Fragilityexamples} See for example: N. Gisin and H. Bechmann-Pasquinucci, Phys. Lett. A {\bf 246}, 1-6
(1998); R. Filip, J. \u Reh\'a\u cek, and M. D\u usek, J. Opt. B:
Quantum Semiclass. Opt. {\bf 3}, 341 (2001); D. Janzing and T.
Beth, Phys. Rev. A {\bf 61}, 052308 (2000).

\bibitem{Jozsa} R. Jozsa, J. Mod. Opt. {\bf 41} (12), 2315 (1994).

\bibitem{Wootters98} W. K. Wootters, Phys.  Rev.  Lett.  {\bf 80}, 2245 (1998).

\bibitem{Bennett96}
C. H. Bennett, D. P. DiVincenzo, J. A. Smolin, and W. K. Wootters,
Phys. Rev. A {\bf 54}, 3824 (1996).

\bibitem{CHSH} J. F. Clauser, M. A. Horne, A. Shimony, and R. A. Holt, Phys. Rev.
Lett. {\bf 23}, 880 (1969).

\bibitem{Werner89}
R. F. Werner, Phys. Rev. A, {\bf 40}, 4277 (1989).

\bibitem{Barrett02}
J. Barrett, Phys. Rev. A, {\bf 65}, 042302 (2002).

\bibitem{Kwi01}
See for example 
P. G. Kwiat, S. Barraza-Lopez, A. Stefanov, and N. Gisin,
Nature, {\bf 409}, 1014 (2001), 
and references therein.

\end{thebibliography}
\end{document}